\documentclass[aps,preprint,showpacs,floatfix]{revtex4}
\usepackage{graphicx,bm,amsmath,dcolumn,}
\begin{document}
\newcommand{\be}{\begin{equation}}
\newcommand{\ee}{\end{equation}}
\newcommand{\half}{\frac{1}{2}}
\newcommand{\ith}{^{(i)}}
\newcommand{\im}{^{(i-1)}}
\newcommand{\gae}
{\,\hbox{\lower0.5ex\hbox{$\sim$}\llap{\raise0.5ex\hbox{$>$}}}\,}
\newcommand{\lae}
{\,\hbox{\lower0.5ex\hbox{$\sim$}\llap{\raise0.5ex\hbox{$<$}}}\,}

\title{Conducting-angle-based percolation in the XY model}
\author{Yancheng Wang$^{1}$,
Wenan Guo$^{1}$,
Bernard Nienhuis$^2$,
and Henk W.J. Bl\"ote$^{3}$ } 
\affiliation{$^{1}$Physics Department, Beijing Normal University,
Beijing 100875, P. R. China}
\affiliation{$^{2}$Instituut voor Theoretische Fysica,
Universiteit van Amsterdam, Valckenierstraat 65, The Netherlands}
\affiliation{$^{3}$ Instituut Lorentz, Leiden University,
P.O. Box 9506, 2300 RA Leiden, The Netherlands}

\date{\today} 
\begin{abstract}
We define a percolation problem on the basis of spin configurations of
the two dimensional XY model. Neighboring spins belong to the same
percolation cluster if their orientations differ less than a certain
threshold called the conducting angle.
The percolation properties of this model are studied % systematically
by means of Monte Carlo simulations and a finite-size scaling analysis.
Our simulations show the existence of percolation transitions when the
conducting angle is varied, and we determine the transition point for
several values of the XY coupling.
It appears that the critical behavior of this percolation model can 
be well described by the standard percolation theory.
The critical exponents of the percolation transitions, as determined by
finite-size scaling, agree with the universality class of the
two-dimensional percolation model on a uniform substrate.
This holds over the whole temperature range,
even in the low-temperature phase where the XY substrate is critical in
the sense that it displays algebraic  decay of  correlations.
\end{abstract}
\pacs{05.50.+q, 64.60.Cn, 64.60.Fr, 75.10.Hk}
\maketitle 

\section{Introduction}

Consider the XY or planar model on the square lattice with periodic
boundary conditions, described by the reduced Hamiltonian
\be
{\mathcal H}=-\frac{J}{k_{\rm B} T} \sum_{<ij>} \vec{s}_i \cdot \vec{s}_j,
\label{Hxy}
\ee
where the sum is over all nearest-neighbor pairs, and the $\vec{s}_i$ are
two-dimensional unit vectors labeled by the site number $i$. We restrict
the nearest-neighbor interaction to be ferromagnetic, i.e., $J > 0$.

When the temperature of this model is lowered, it undergoes a phase
transition of an interesting character, which was explained by
Kosterlitz and Thouless \cite{KT}. More exact results for the exponents
were obtained by Nienhuis \cite{Nhon} for an O(2) model in the same
universality class. These results show that the renormalization
exponent $y_t$ of the temperature is equal to 0, which means that
the temperature-driven transition is of infinite order, i.e. the
specific-heat singularity is extremely weak. In contrast, the magnetic
susceptibility displays a very strong divergence when the temperature is
lowered to the transition point.

At temperatures $T>0$ below the transition point there is no
spontaneous long-range order in the sense that the magnetization is
zero \cite{MW}. Instead, the low-temperature phase resembles a critical
state; the correlations decay algebraically, with exponents that are
still dependent on the temperature.

Recently, there has been ample attention to percolation problems defined
on a critical substrate, see e.g. Refs.~\onlinecite{XYH,DBN,DDGQB,DDGB}
and references therein. It thus appears that, for Potts and O($n$) models,
the universal properties of such percolation transitions do reflect 
the nature of the critical substrate. These investigations are based
on model representations with discrete degrees of freedom.

It is thus interesting to investigate related problems using a substrate 
with continuous degrees of freedom. For instance, the mechanical properties
of static granular matter can be analyzed in terms of the so-called force
networks \cite{GHM,JNB}. By introducing a threshold force, such that
forces exceeding the threshold form clusters, a percolation transition
is seen at a critical force threshold. 
This approach was applied to different models for granular piles,
that are expected to belong to different universality classes. 
It was found that also the corresponding percolation behavior could
discriminate between the different models \cite{OSN}. This might suggest
that such substrate dependence is a general phenomenon for critical
models with continuous degrees of freedom, i.e., the critical behavior
of the percolation clusters might generally reflect the long-range
correlations of the original degrees of freedom.

In order to shed more light on this issue, the present work investigates
a percolation problem using the substrate of the two-dimensional XY-model.
In particular we are interested how the universal properties of the
percolation transition will depend on the temperature of the underlying
XY model. We define the percolation problem such as to  depend only on
the XY configuration, and not on any additional random variables.
This is achieved by introducing a ``conducting angle'' $\theta$ such
that neighboring spins whose orientations differ by less than $\theta$ are
connected by a percolation bond. This name is based on the analogy
with a conductance problem of conducting units on a lattice, such that
neighboring units are in electrical contact only if their orientations
match to a sufficient degree. 

\begin{figure}[htbp]
\includegraphics[scale=0.7]{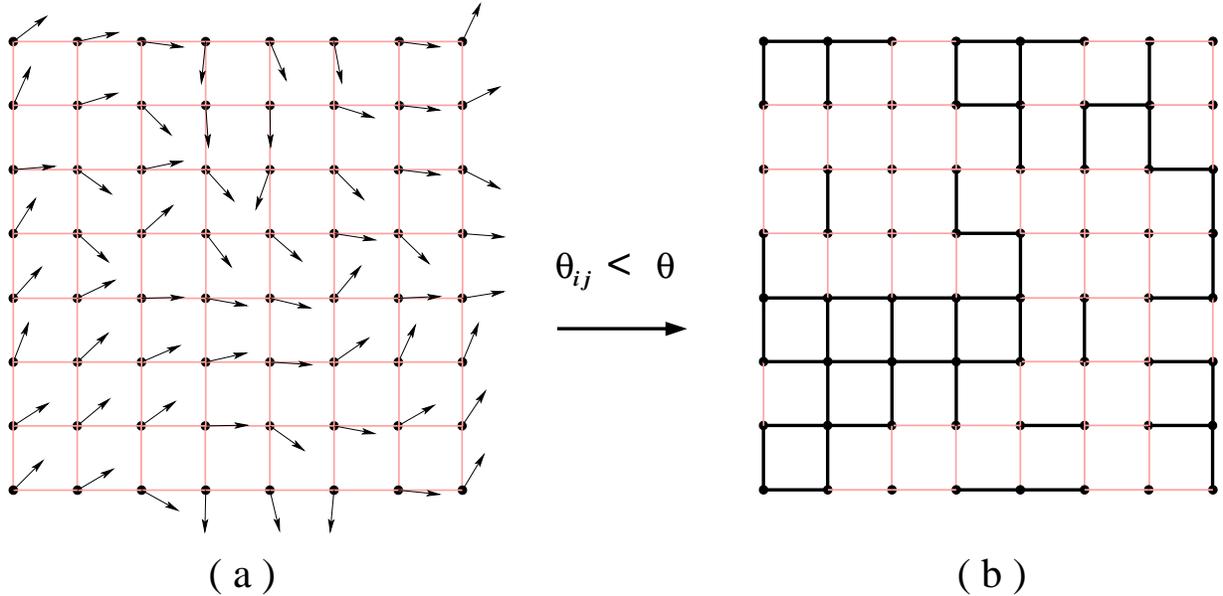}
\centering
\caption{ Construction of a bond-percolation configuration (b) from an
XY spin configuration (a). If the angle $\theta_{ij}$ between a pair (i, j)
of neighboring spins is less than a given angle $\theta$, the neighboring
spins are connected by a bond, so that the lattice decomposes into a 
system of percolation clusters. The threshold was chosen as $\theta=\pi/6$ 
in this figure.}
\label{xy-network}
\end{figure}

It is clear that this conducting angle also defines a threshold pair
energy, below which a pair of neighboring XY spins is connected by a
percolation bond. These percolation bonds define a bond percolation 
configuration involving a complete decomposition of the lattice in
percolation clusters.
An example of an XY configuration with the corresponding percolation
cluster decomposition is shown in Fig.~\ref{xy-network}.
It is obvious that the resulting bond percolation configurations will
not percolate for $\theta=0$, and that they will percolate for
$\theta=\pi$ and larger. We are thus left with the task to find the
percolation threshold $\theta_{\rm c}$ and the critical exponents as
a function of the temperature.
To answer these questions, we perform simulations of the two-dimensional
(2D) XY spin model, which are described in Section \ref{algo}.
Section \ref{numres} presents the numerical results, including the
critical parameters. We include a short discussion in Sec.~\ref{disc}.

\section{Algorithm}
\label{algo}
For reasons of efficiency, we make use of a cluster algorithm \cite{SW}
to simulate the XY model. We applied the single-cluster algorithm 
formulated by  Wolff \cite{W2} to the model on the square lattice.
We recall the steps involved in one Wolff cluster flip:
\begin{enumerate}
\item Choose an arbitrary direction as the $y$-direction, and denote 
by $\alpha_i$ the angle between the spin $\vec{s_i}$ and the $y$-axis.
Define an Ising spin $s_i=\pm 1$ with the same sign as the $y$-component
of $\vec{s_i}$. Thus, the nearest-neighbor interaction term in
${\mathcal H}_{ij}$ between spins $i$ and $j$ reads 
\be
{\mathcal H}_{ij}=
-K \vec{s}_i\cdot \vec{s}_j=-K s_i^x s_j^x-Ks_i^y s_j^y
     =-K s_i^x s_j^x-K \cos(\alpha_i)\cos(\alpha_j)\, s_i s_j \, ,
\ee
where $K\equiv J/k_{\rm B}T$.
As far as the dependence of ${\mathcal H}_{ij}$ on the Ising variables
${s_i}$ and ${s_j}$ is concerned, this is an Ising coupling between the
two spins with strength $K_{ij}=K\cos \alpha_i \cos \alpha_j$.
One may update the Ising variables using a cluster algorithm that 
takes into account these position-dependent couplings.
The following steps are used to update the $y$-components of the spins. 

\item Choose a spin randomly, say on site $i$. 
For each nearest-neighbor site $j$ of $i$, connect $i$ and $j$ by a 
bond with probability $p_{ij}=\max (0,1-e^{-2K_{ij}})$. 
Then do the same for each of the nearest-neighbor sites of each
newly connected site, and so on. 
The process continues until no more new sites are connected. Then, the
construction of the cluster, which contains all sites connected via
some path of bonds to site $i$, is finished. 

\item Change the sign of the $y$-components of all spins in that cluster.

\end{enumerate}
\section{Numerical results and analysis} 
\label{numres}
During the simulations, we constructed percolation cluster decompositions
on the basis of a chosen conducting angle $\theta$.
For each such decomposition we sampled several quantities in order to
estimate the second moment of the cluster size distribution $S_2$, 
the probability $P$ that there exists a cluster that wraps the system,
a dimensionless ratio $Q$ that can be related to the Binder cumulant,
and the cluster size distribution function $n_s$. We shall proceed to 
introduce these quantities in more detail and list their expected
finite-size scaling properties.

Since we are using finite systems with periodic boundary conditions,
we define a "wrapping cluster" \cite{Pinson} as a percolation cluster
that connects to
itself along at least one of the periodic directions. For each  XY
configuration $S$ we thus define a quantity $p(S,\theta)$ that has the
value 1 if there exists a wrapping cluster, and $p(S,\theta)=0$ otherwise.
Thus, for a system with finite size $L$, the probability $P(K,\theta,L)$
that a wrapping cluster exists is given by 
\be
P(K,\theta,L)= \langle p(S,\theta)\rangle \, ,   
\ee
where the ensemble average is taken for an XY system with linear size
$L$ and coupling $K$.
If there is a percolation transition at a conducting angle $\theta_{\rm c}$,
one expects that $\theta$ plays the role of a temperature-like 
variable, and thus that the finite-size scaling behavior \cite{FSS}
of $P$ is described by
\begin{equation}
P(K,\theta,L)=P^{(c)}+a(\theta-\theta_{\rm c})L^{y_{_p}}+\cdots+
b_1L^{y_{_1}}+b_2L^{y_{_2}}+\cdots,
\label{psfit}
\end{equation}
where $y_p$ controls the scaling of $\theta$ and acts as a 
temperature-like exponent, corresponding with the bond dilution exponent
in the language of the percolation model. The exponents $y_1$,
$y_2$, etc.~are correction-to-scaling exponents, which are unknown in
principle. The constant $P^{(c)}$, which is defined as the value of $P$
at $\theta_{\rm c}$ in the limit $L \to \infty$, and the exponent $y_p$
are universal, but the universality class of the present percolation
problem remains to be determined.

The second moment $S_2$ of the percolation cluster size distribution 
can also be viewed as the mean size of the cluster containing an arbitrary
point. It is defined as 
\begin{equation}
S_2=\frac{1}{N^{2}} \left< \sum_{i}^{N_c}s_i^2\right> \, ,
\label{C2def}
\end{equation} 
where $s_i$ is the size of the $i$-th cluster, $N_c$ the total number
of clusters for a configuration, and $N=L^2$ is the volume of the system.
The quantity $S_2$ is also closely related with the generalization
$\chi_{\rm RC}$ of the $q$-state Potts magnetic susceptibility to the
random-cluster model, applied to the special case $q=1$.
This relation is expressed as $\chi_{\rm RC}=N S_2$.

In analogy with the Potts susceptibility, we expect the following
finite-size-scaling behavior for $S_2$ in the neighborhood of a
percolation threshold at $\theta_{\rm c}$:
\begin{eqnarray}
S_2&=&L^{2y_{_h}-2d}[a_0+a_1(\theta-\theta_{\rm c})L^{y_{_p}}+\cdots+
b_1L^{y_{_1}}+b_2L^{y_{_2}}+\cdots] \, ,\label{C2fs0}
\end{eqnarray}
where $y_h$ is the fractal dimension of critical percolation clusters.
At the percolation threshold $\theta_{\rm c}$, this equation reduces to
\begin{eqnarray}
S_2&=&L^{2y_{_h}-2d}(a_0+b_1 L^{y_{_1}}+b_2 L^{y_{_2}}+\cdots) \, .
\label{C2fs}
\end{eqnarray}

We also define a dimensionless ratio related to the Binder cumulant
\cite{Binder} as
\begin{equation}
Q=\frac{S_2^2}{3S_2^{(2)} - 2S_4} \, ,
\label{Q-def}
\end{equation}
where $S_2^{(2)}$ is defined as
\begin{equation}
S_2^{(2)} = \frac{1}{N^{4}}\left<\left(\sum_{i}^{N_c}s_i^2\right)^2 \right>,
\label{S22def}
\end{equation}
and $S_4$ as
\begin{equation}
S_4=\frac{1}{N^{4}}\left< \sum_{i}^{N_c}s_i^4\right> \, .
\label{S4def}
\end{equation} 
The relation with the Binder cumulant is based on the fact that
Eq.~(\ref{Q-def}) is obtained when the Binder ratio
$\langle m^2\rangle^2/\langle m^4 \rangle $ of magnetization moments of the
Ising model is expressed in the language of the $q=2$ random-cluster model.
In analogy with the Binder cumulant, we expect that, in the neighborhood 
of the percolation threshold, it behaves like 
\begin{equation}
Q(K,\theta ,L)=Q^{(c)}+a_1 (\theta-\theta_{\rm c})L^{y_p}
+\cdots+b_1L^{y_{_1}}+b_2L^{y_{_2}}+\cdots \, . 
\label{Qfit}
\end{equation}

\subsection{Wrapping probability as a function of $\theta$}
We simulated XY systems with linear sizes $L=8$, 16, 32, 64, 128, and 256
at different inverse temperatures: $K=0$, 0.001, 0.01, 0.10, $\cdots$,
5.00, and 10.00. For $K=0.9$ and $K=1.0$, system sizes up to
$L=1024$ and $2048$ were simulated respectively. 
Typically, $10^7$ XY configurations were sampled. The samples were taken 
at intervals consisting of one  Metropolis sweep and four Wolff clusters.
For each of these XY configurations, percolation cluster decompositions
were constructed with several different values of the conducting angle.

\begin{figure}
\includegraphics[scale=0.95]{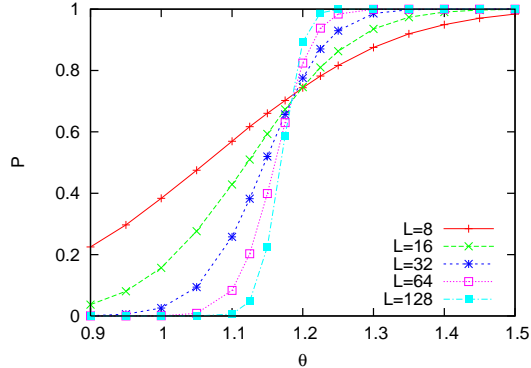}
\centering
\centering
\caption{(color online). Wrapping probability versus $\theta$ at $K=0.5$.
 The curves are added as a guide for the eye.}
\label{psn1}
\end{figure}

\begin{figure}
\includegraphics[scale=0.95]{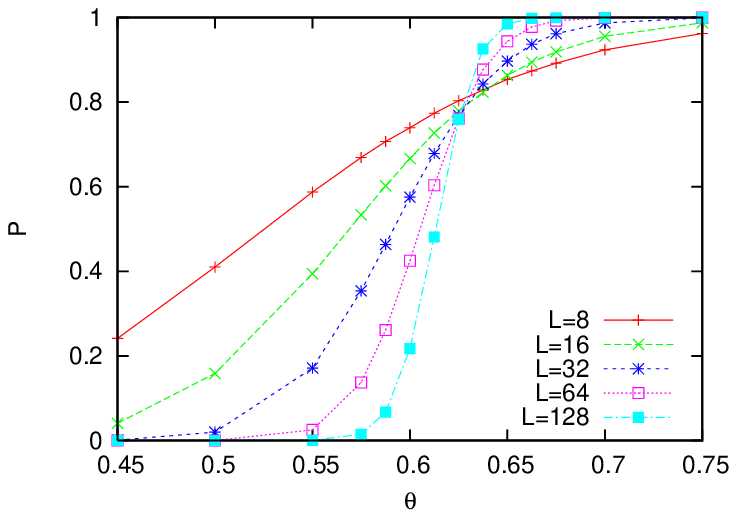}
\includegraphics[scale=0.95]{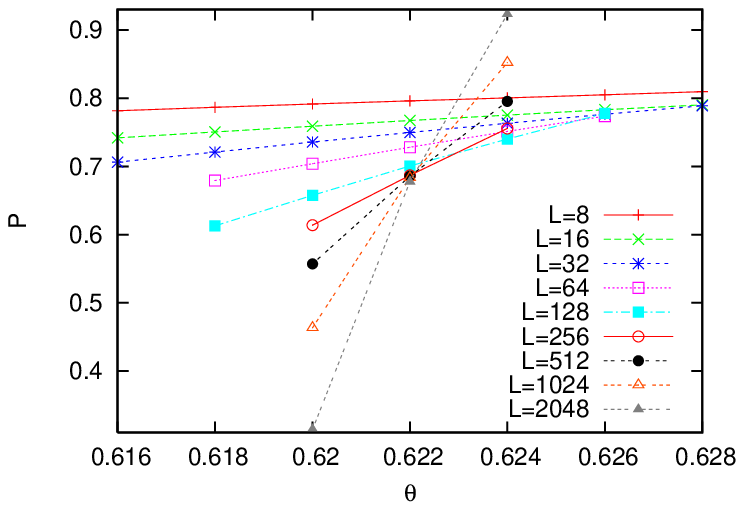}
\centering
\caption{(color online). Wrapping probability versus $\theta$ at $K=1$.
The right-hand figure shows the details in the vicinity of $\theta_{\rm c}$,
where the finite size effect becomes relatively strong. 
The curves are added as a guide for the eye.}
\label{psn2}
\end{figure}

\begin{figure}
\includegraphics[scale=0.95]{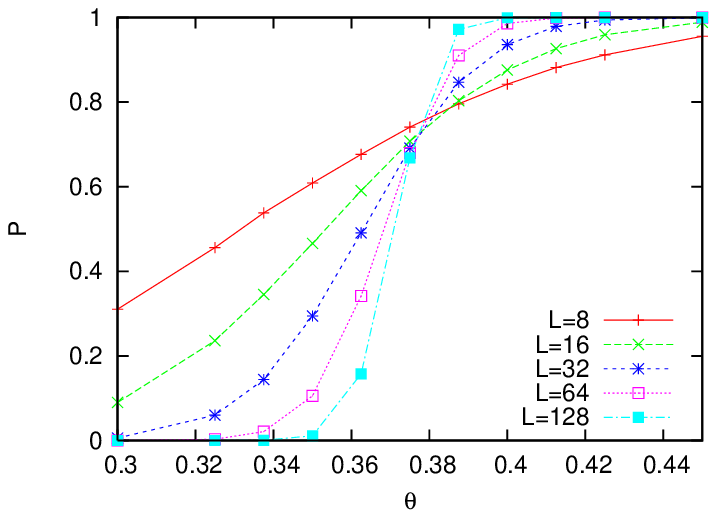}
\centering
\caption{(color online). Wrapping probability versus $\theta$ at $K=2.0$.
 The curves are added as a guide for the eye.}
\label{psn3}
\end{figure}

\begin{figure}[htb]
\includegraphics[scale=1]{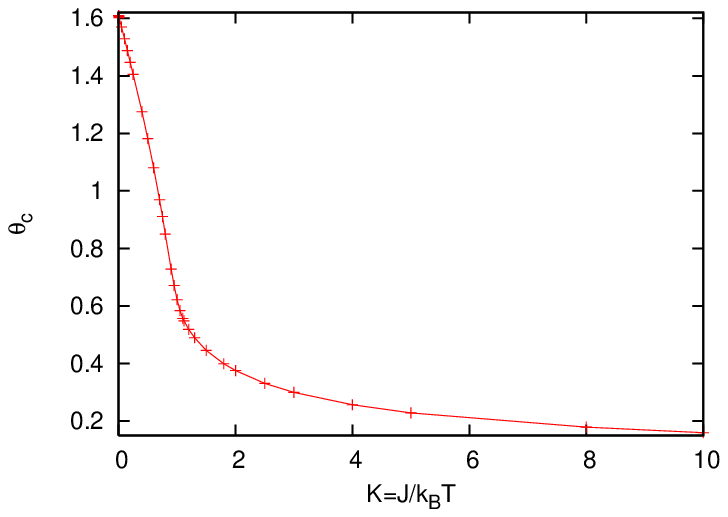}
\centering
\caption{ Percolation threshold $\theta_{\rm c}$ versus $K$. 
For $K=0$, the
critical value of the conducting angle is 1.61078 radians or about
92 degrees. This value applies to a system of randomly oriented
conductors on the square lattice. The curve is added as a guide for the
eye, and estimated error bars are smaller than the size of the symbols.}
\label{phase}
\end{figure}

Some results for the wrapping probability $P(K,\theta,L)$ as a function
of the conducting angle are shown in Figs.~\ref{psn1}-\ref{psn3}.
These figures show clear intersections corresponding with percolation
thresholds. Furthermore, the behavior appears to be remarkably similar
for different values of the XY coupling $K$.
We applied the least-squares method to fit the wrapping probability by 
the finite-size-scaling equation (\ref{psfit}).
The results for the  percolation threshold $\theta_{\rm c}$ are
shown in Fig.~\ref{phase} as a function of the XY coupling $K$.
They are also included in Table \ref{table1} for each value of $K$,
together with the results for the exponent $y_p$ and the universal
probability $P^{(c)}$. The results for the latter quantity reproduce,
within error bounds, the literature value $P^{(c)}=0.690473725$ 
which applies to the ordinary two-dimensional percolation
model \cite{Pinson,ZLK,NZ}.
\bigskip 

\subsection{Dimensionless ratio $Q$ as a function of $\theta$}

\begin{figure}[htbp]
\includegraphics[scale=0.95]{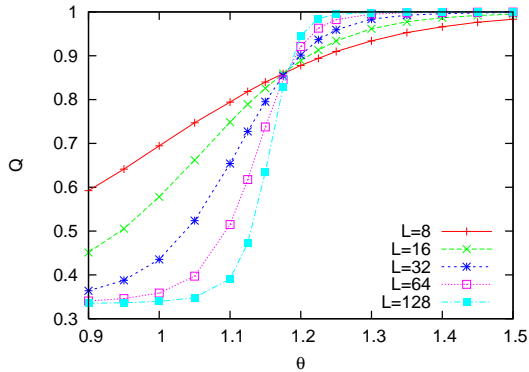}
\centering
\caption{(color online). Dimensionless ratio $Q$ versus conducting angle
$\theta$ for $K=0.5$. The curves are added as a guide for the eye. }
\label{Q-0.5}
\end{figure}

\begin{figure}[htbp]
\includegraphics[scale=0.95]{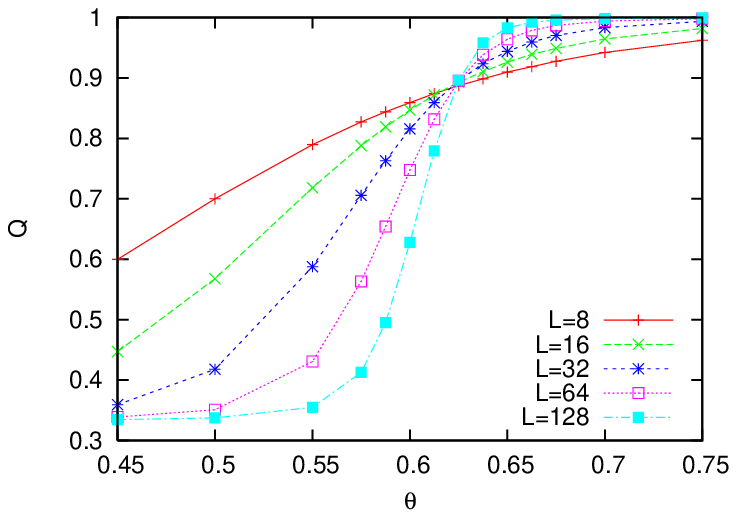}
\includegraphics[scale=0.95]{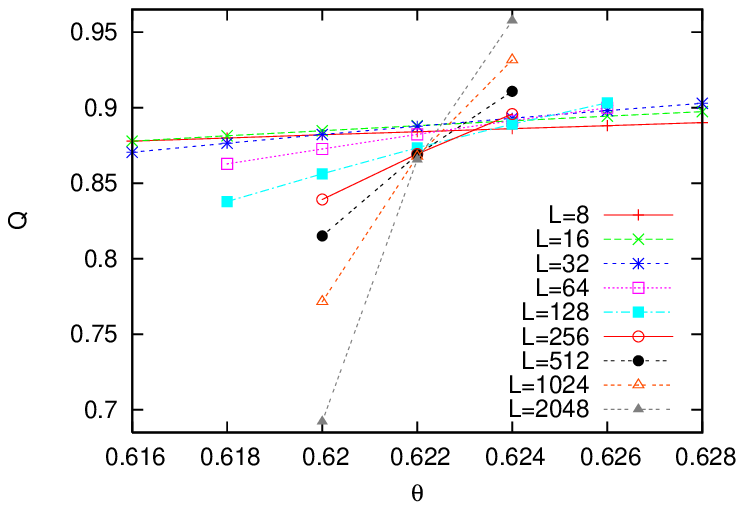}
\centering
\caption{(color online). Dimensionless ratio $Q$ versus conducting angle 
$\theta$ for $K=1.0$. The right-hand figure shows the details in the
vicinity of $\theta_{\rm c}$. The curves are added as a guide for the eye.}
\label{Q-1.0}
\end{figure}

\begin{figure}[htbp]
\includegraphics[scale=0.95]{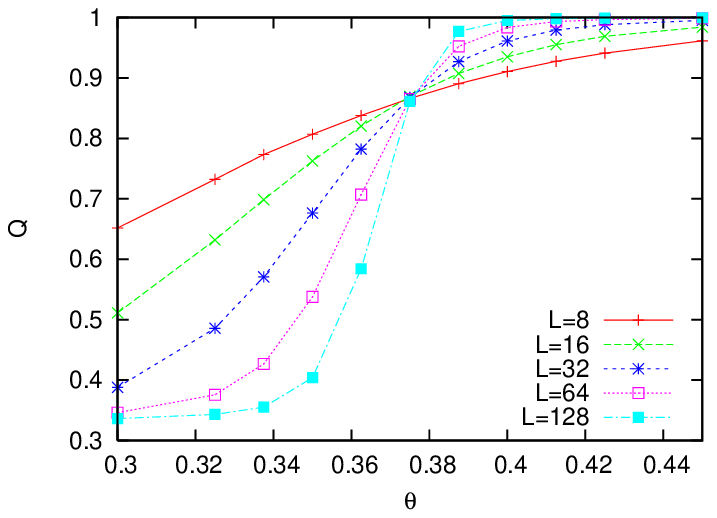}
\centering
\caption{(color online). Dimensionless ratio $Q$ versus conducting angle 
$\theta$ for $K=2.0$. The curves are added as a guide for the eye.}
\label{Q-2.0}
\end{figure}

We also sampled the ratio $Q$ for systems at different inverse
temperatures. Some of the data are shown in Figs.~\ref{Q-0.5} to 
\ref{Q-2.0}. They display the same general behavior as the 
wrapping probability in the preceding subsection.

A least-squares analysis of the data for $Q(K,\theta,L)$ on the basis of
the finite-size-scaling equation (\ref{Qfit}) results in the estimates
of $y_p$, $Q^{(c)}$, and the percolation threshold $\theta_{\rm c}$.
The estimates of $y_p$ and $\theta_{\rm c}$  are in agreement with
those found by fitting $P(K,\theta,L)$. The estimates of the universal
value $Q^{(c)}$ are also listed in Table \ref{table1} for each $K$.
Their values are in agreement with the literature value \cite{DB}
$0.87053~(2)$ for the ordinary two-dimensional percolation model.

\subsection{The second moment $S_2$ at the percolation threshold}

We simulated the model with system sizes  $L=8$, 16, 32, 64, 128, 256 and
$512$ at the estimated percolation threshold $\theta_{\rm c}$ for each $K$,
and sampled the second moment $S_2$ of the cluster-size distribution.
Samples were taken at intervals consisting of one  Metropolis sweep
and four Wolff clusters.
The numbers of samples taken for system sizes $L=8$ to $32$, $L=64$ to $128$,
and $L=256$ to $512$ are $10^7$, $4 \times 10^6$ and $2 \times 10^6$
respectively. For $K=1.0$,
which is in the vicinity of the Kosterlitz-Thouless (KT) transition,
additional simulations took place for system sizes $L=1024$ and 2048,
involving several times $10^6$ samples per system size.

\begin{figure}
\includegraphics[scale=1.0]{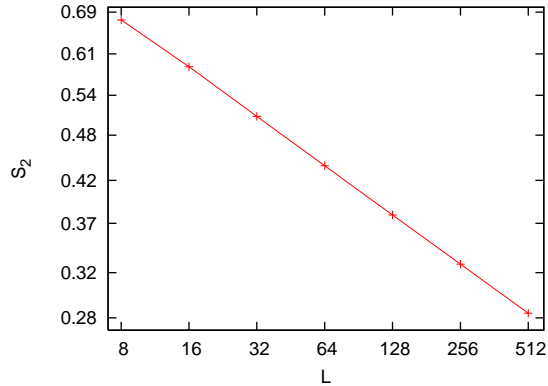}
\centering
\caption{ Second moment $S_2$ of the cluster size
distribution versus finite size $L$, for XY coupling $K=0.5$.
The curve is added as a guide for the eye, and 
estimated error bars are smaller than the size of the symbols.}
\label{b0.001-0.01}
\end{figure}

\begin{figure}
\includegraphics[scale=1.0]{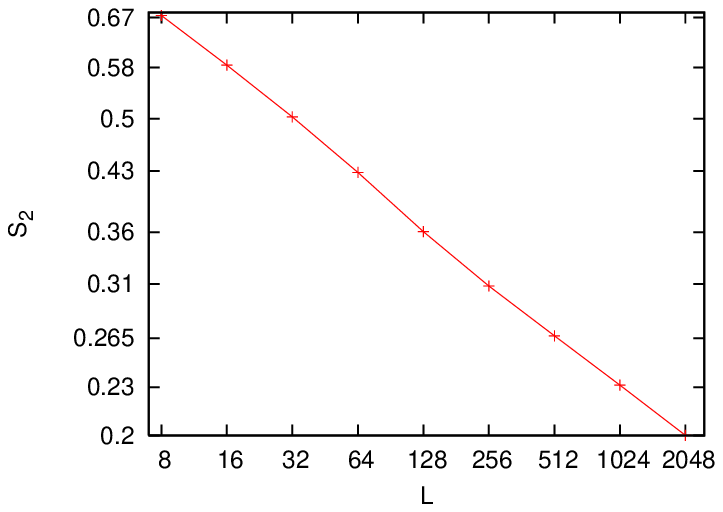}
\centering
\caption{ Second moment $S_2$ of the cluster size
distribution versus finite size $L$, for XY coupling $K=1.0$.
The curve is added as a guide for the eye, and
estimated error bars are smaller than the size of the symbols. }
\label{b0.1-0.5}
\end{figure}

\begin{figure}
\includegraphics[scale=1.0]{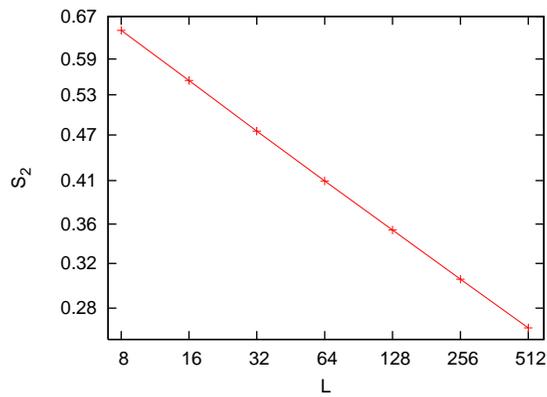}
\centering
\caption{ Second moment $S_2$ of the cluster size
distribution versus finite size $L$, for XY coupling $K=2.0$.
The curve is added as a guide for the eye, and
estimated error bars are smaller than the size of the symbols. }
\label{b0.7-0.8}
\end{figure}

Some of the data for $S_2$ are shown as a function of the system size 
$L$ in Figs.~\ref{b0.001-0.01}-\ref{b0.7-0.8} for some values of $K$.
These figures use logarithmic scales, so that linear behavior means that
$S_2$ behaves as a power of $L$, in accordance with criticality.
For $K=1.0$, which is close to the KT transitions, small deviations
from linearity are visible.

We fitted $S_2$ by  the finite-size-scaling formula Eq.~(\ref{C2fs}),
and thus obtained estimates of the fractal dimension $y_h$,
which are listed in Table \ref{table1} for several values of $K$.  For
$K=1.0$, a satisfactory fit could only be obtained by discarding
system sizes $L<128$ and including system sizes up to $L=2048$.
\subsection{The distribution $n_s$ of the cluster size $s$}
At the percolation transition $\theta_{\rm c}$, the requirement that the cluster
size distribution scales in a covariant way yields a power law for this
distribution
\begin{equation}
n_s (s,\theta_{\rm c}) \propto s^{-\tau} \, ,
\label{ns2-2}
\end{equation}
where $\tau$ is a critical exponent equal to $1+d/y_h$, with
$d=2$ dimensions and the fractal dimension $y_h$ of the percolation
clusters. For the ordinary $d=2$ percolation model, one has 
$y_h=91/48$ \cite{CG,Cardy}, and thus $\tau=187/91$.

\begin{figure}
\includegraphics[scale=1.0]{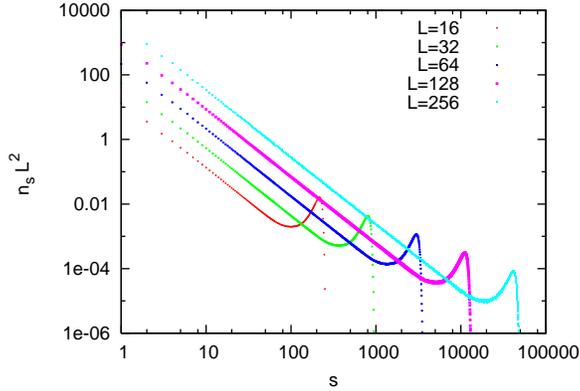}
\centering
\caption{(color online). Cluster size distribution function as a 
function of cluster size $s$ for $K=0.5$, $\theta_{\rm c} = 1.18179$.}
\label{ns-0.5}
\end{figure}

\begin{figure}
\includegraphics[scale=1.0]{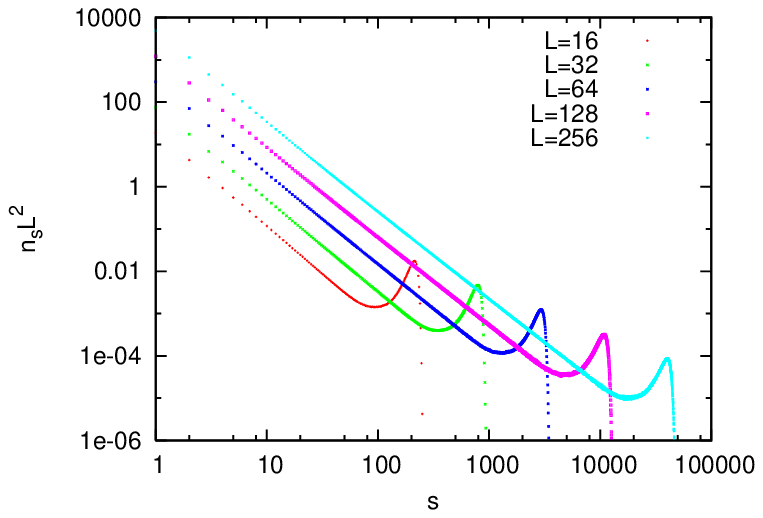}
\centering
\caption{(color online). Cluster size distribution function as a
function of cluster size $s$ for $K=1.0$, $\theta_{\rm c} = 0.62210$.}
\label{ns-1.0}
\end{figure}

\begin{figure}
\includegraphics[scale=1.0]{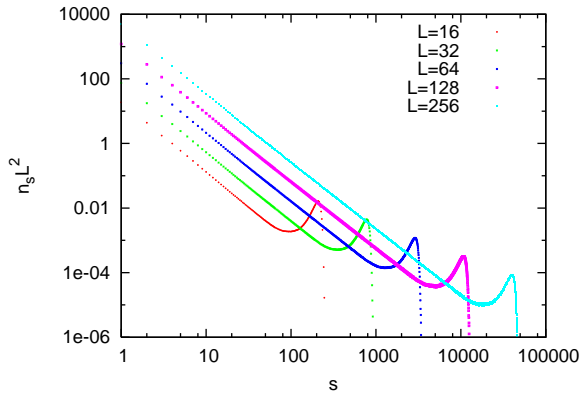}
\centering
\caption{(color online). Cluster size distribution function as a
function of cluster size $s$ for $K=2.0$, $\theta_{\rm c} = 0.37560$.}
\label{ns-2.0}
\end{figure}

Parts of the data for $n_s$ are shown in Figs.~\ref{ns-0.5}-\ref{ns-2.0} 
as a function of cluster size $s$, for different values of the XY
coupling $K$.
The axes in these figures use logarithmic scales. The data points
are, in a wide range of cluster sizes, well approximated by straight
lines, corresponding with a scale-invariant distribution according to
a power law. The slopes of these lines are close to the value
of $\tau$ that applies to the ordinary percolation model.
The curves in Figs.~\ref{ns-0.5}-\ref{ns-2.0} can obviously be brought
to an approximate data collapse by introducing suitable prefactors,
but it appears that there is still an appreciable finite-size
dependence.

For a numerical determination of $\tau$ from the data for $n_s$, we
use a least-squares analysis of the range of $s$ where $n_s$ is
almost linear in Figs.~\ref{ns-0.5}-\ref{ns-2.0}, thus excluding
small clusters smaller than 20, and the largest clusters exceeding
a size $L^{y_h}/10$.
%Henk, we change 20-->10.
Even then, two correction terms had to be included in order to obtain
satisfactory residuals, according to the fit formula
\begin{equation}
n_s(s,\theta_{\rm c},L)=a_0 s^{-\tau}(1+a_1 s^{y_1}+a_2 L^{y_2}) \, ,
\label{nsfit}
\end{equation}
where the term with coefficient $a_2$ is significant only for $K>1$.
Satisfactory fits were obtained for all inverse temperatures. For example,  
for $K=0.75$, we fitted the data simultaneously for system sizes $L=32$,
64,$\cdots$, 512 using the above formula, and found $\tau=2.055~(2)$, 
$y_1=-0.57~(1)$ with $\chi^2$ per degree of freedom almost equal to 1. 
For $K=2.0$, a fit of the data for system sizes  $L=32$,64,$\cdots$, 512 
yielded $\tau=2.054~(2)$, $y_1=-1.0~(1), y_2=-1.6~(1)$, again with 
a satisfactory value of $\chi^2$  as compared with the number of 
degrees of freedom.

We thus obtained estimates of the exponent $\tau$ that agree well
with the value $\tau=187/91=2.0549\cdots$ for the ordinary 2D
percolation model, for several temperatures of the XY model.

\begin{table}
\caption{Numerical results for the percolation threshold $\theta_{\rm c}$,
the bond dilution exponent $y_p$, the fractal dimension $y_h$, the
universal wrapping probability $P^{(c)}$ and the universal dimensionless
ratio $Q^{(c)}$. The exact values for two-dimensional percolation 
exponents are $ y_p=3/4$ and $ y_h=91/48=1.895833\cdots$.}
\begin{tabular}{|c|c|c|c|c|c|}
\hline
\hline
$K $  & $\theta_{\rm c}$  & $ y_p $  & $ y_h $    & $P^{(c)}$  & $Q^{(c)}$ \\
\hline
0     & 1.61078 (2) & 0.750(2) & 1.8961 (3) & 0.6906 (2) & 0.8708 (3)\\
0.001 & 1.60994 (2) & 0.748(3) & 1.8961 (3) & 0.6908 (3) & 0.8707 (5)\\
0.010 & 1.60252 (2) & 0.750(2) & 1.8959 (3) & 0.6905 (3) & 0.8706 (3)\\
0.100 & 1.52866 (2) & 0.751(2) & 1.8959 (3) & 0.6905 (3) & 0.8708 (5)\\
0.200 & 1.44651 (2) & 0.752(2) & 1.8958 (3) & 0.6903 (4) & 0.8708 (5)\\
0.500 & 1.18179 (2) & 0.750(2) & 1.8956 (3) & 0.6903 (3) & 0.8709 (5)\\
0.750 & 0.91074 (2) & 0.751(2) & 1.8957 (3) & 0.6901 (5) & 0.8704 (5)\\
0.800 & 0.85002 (2) & 0.749(5) & 1.8957 (3) & 0.6905 (3) & 0.8702 (5)\\
0.900 & 0.72804 (2) & 0.751(4) & 1.8958 (3) & 0.6902 (4) & 0.8705 (5)\\
1.000 & 0.62209 (2) & 0.74 (2) & 1.894~ (2) & 0.6907 (4) & 0.8705 (5)\\
1.100 & 0.55695 (1) & 0.753(3) & 1.8954 (5) & 0.6902 (6) & 0.8709 (5)\\
1.120 & 0.54815 (1) & 0.752(3) & 1.8958 (2) & 0.690~ (1) & 0.8706 (5)\\
1.200 & 0.51869 (1) & 0.751(2) & 1.8959 (2) & 0.6903 (5) & 0.8707 (5)\\
1.500 & 0.44598 (2) & 0.753(3) & 1.897~ (2) & 0.6907 (3) & 0.8708 (5)\\
2.000 & 0.37560 (2) & 0.752(3) & 1.895~ (1) & 0.6900 (8) & 0.8708 (3)\\
5.000 & 0.22851 (1) & 0.751(2) & 1.8958 (3) & 0.690~ (1) & 0.8708 (5)\\
10.00 & 0.15981 (2) & 0.751(4) & 1.895~ (1) & 0.689~ (3) & 0.8708 (5)\\
\hline
\hline
\end{tabular}
\label{table1}
\end{table}

\section{Discussion}
\label{disc}
The numerical results presented in Sec.~\ref{numres} for the
exponents $\tau$, $y_h$ and $y_p$, as well as for the universal
probability $P^{(c)}$ and the universal ratio $Q^{(c)}$, agree
accurately with the literature values applying to the two-dimensional
percolation model. Thus our analysis shows the existence
of a transition in the ordinary percolation universality class,
independent of the XY temperature. This result is as expected in the 
high-temperature phase where the XY spins display strong random
disorder, but may seem somewhat surprising in the critical region
including the low-temperature range. 
The problem discussed here can be viewed as a correlated percolation
problem, in which the bond probabilities correlate as the nearest-neighbor 
differences in the XY model. For such long-range correlated percolation 
Weinrib \cite{weinrib} formulated a generalized Harris criterion to decide 
if the correlations are relevant for the percolation behavior or not.
The critical behavior is expected to be in the universality class of 
ordinary (uncorrelated) percolation if $a \nu > 2$ for correlations decaying 
with distance as $r^{-a}$, with $\nu$ the percolation correlation length 
exponent $\nu=4/3$. Indeed in the case at hand $a=2$, and the correlations 
should thus be irrelevant.
It stands in a strong
contrast with a recent analysis \cite{DDGB} of a percolation problem
defined on the basis of the two-dimensional O($n$) model where $n$ is
a continuously variable parameter. For the O(2) model, which belongs to
the same universality class as the XY model, it was found \cite{DDGB}
that the percolation transition was driven by a bond dilution field
that is only marginally relevant, i.e., $y_p=0$ at that transition,
while $y_p=3/4$ for ordinary percolation. It thus appears that the
character of a percolation transition on a critical substrate depends
on the precise definition of the percolation problem. The percolation
problem of Ref. \cite{DDGB} was defined within regions separated by
loops as defined in the context of the O($n$) loop model. These loops
obviously display fractal properties\cite{DDGQB} at O($n$) criticality,
and thus affect the nature of the percolation transition. In contrast,
the spatial variation of the angle between neighboring XY spins is very
smooth near criticality, and does not display boundary-like structures,
even at the KT transition. While it is natural that the KT transition
is reflected in some way in the behavior of the percolation transition
at the critical value of the conducting angle, such effects are exposed
by our numerical results only in terms of slow convergence  in the
analysis of the finite-size data.

\acknowledgments
W. G. acknowledges hospitality extended to him by the Lorentz Institute.
This work is supported by the Lorentz Fund, by the NSFC under Grant No.
10675021, and by the HSCC (High Performance Scientific Computing Center)
of the Beijing Normal University.

\end{document}